# metaSPAdes: a new versatile metagenomics assembler

Sergey Nurk[1,*,**], Dmitry Meleshko[1,*], Anton Korobeynikov[1,2] and Pavel A Pevzner[1,3]

[1]Center for Algorithmic Biotechnology, Institute for Translational Biomedicine,
St. Petersburg State University, St. Petersburg, Russia

[2]Department of Statistical Modelling, St. Petersburg State University, St. Petersburg, Russia

[3]Department of Computer Science and Engineering, University of California,
San Diego, USA

*These authors contributed equally to this work
**corresponding author, sergeynurk@gmail.com

## Abstract

While metagenomics has emerged as a technology of choice for analyzing bacterial populations, assembly of metagenomic data remains challenging thus stifling biological discoveries. Moreover, recent studies revealed that complex bacterial populations may be composed from dozens of related strains thus further amplifying the challenge metagenomics assembly. metaSPAdes addresses various challenges of metagenomics assembly by capitalizing on computational ideas that proved to be useful in assemblies of single cells and highly polymorphic diploid genomes. We benchmark metaSPAdes against other state-of-the-art metagenome assemblers and demonstrate that it results in high-quality assemblies across diverse datasets.

## Introduction

Metagenome sequencing has emerged as a technology of choice for analyzing bacterial populations and discovery of novel organisms and genes (Venter et al. 2004; Tyson et al. 2004; Yooseph et al. 2007; Arumugam et al. 2011). In one of the early metagenomics studies, Venter et al. (2004) attempted to assemble the complex Sargasso Sea microbial community but, as the paper stated, failed. On the other side of the spectrum of metagenomics studies, Tyson et al. (2004) succeeded in assembling a simple metagenomic community from a biofilm consisting of a few species.

These landmark studies used conventional computational assembly tools with minor modifications. Since they were published, many specialized metagenomics assemblers have been developed (Laserson et al. 2011; Peng et al. 2011; Koren et al. 2011; Peng et al. 2012; Namiki et al. 2012; Boisvert et al. 2012; Haider et al. 2014; Li et al. 2016). However, bioinformaticians are still struggling to bridge the gap between assembling simple and complex metagenomics communities (see Gevers et al. (2012) for a review). Meanwhile, many researchers succeeded in reconstructing individual abundant genomes out of complex metagenomes (Dupont et al. 2012; Iverson et al. 2012; Hess et al. 2011;

Dupont et al. 2016) by performing *de novo* assembly followed by a partition of contigs into *bins* based on coverage depth, sequence composition, mate-pair information, and other criteria (Dick et al. 2009; Wu and Ye 2011; Wu et al. 2014). However, this approach often faces difficulties because of high fragmentation of metagenomic assemblies, which negatively affects both the accuracy of binning and the contiguity of genomes attributed to specific bins.

Recent applications of single-cell (Kashtan et al. 2014) and *TruSeq Synthetic Long Reads* (Sharon et al. 2015) technologies revealed an enormous microdiversity of related strains within various microbial communities. While strains share most of the genomic sequence, they often have significant variations arising from mutation, mobile element insertion, horizontal gene transfer, or genome rearrangement. For example, single-cell sequencing revealed that *Prochlorococcus* (the most abundant photosynthetic bacteria on Earth) can be viewed as a "federation" of diverse cells representing combinatorial arrangements of alleles with some subpopulations separated by millions of years of evolution (Kashtan et al. 2014, Biller et al., 2015). Using TruSeq Synthetic Long Reads (TSLRs), Sharon et al. (2015) showed that the most abundant species in their soil samples are represented by dozens of related strains. Moreover, authors argued that this microdiversity was responsible for the overly-fragmented reconstruction of the corresponding genomes from short-read libraries.

Assembling bacterial communities with high level of microdiversity is just one of key metagenomics challenges that we list below:

*Non-uniform read coverage of various species within a metagenome.* Widely different abundance levels of various species in a microbial sample result in a highly non-uniform read coverage across different genomes. Moreover, the read coverage of most species is considerably lower than the coverage in typical assembly projects of cultivated genomes. As the result, standard assembly techniques aimed at isolate genomes with rather uniform coverage, generate more fragmented and error-prone metagenomics assemblies.

*Similarities between different bacterial species.* Genomes of various species within a microbial community often share highly conserved regions. Besides complicating the assembly and fragmenting contigs, such "interspecies repeats", together with low coverage of most species, may trigger intergenomic assembly errors.

*Differences between closely related strains of the same bacterial species.* Many bacterial species in a microbial sample are represented by *strain mixtures*, i.e. multiple related strains with varying abundances (Kashtan et al. 2014; Biller et al. 2014; Rosen et al. 2015; Sharon et al. 2015). Although various studies outside the field of metagenomics extensively addressed the similar challenge of assembling *two* haplomes within a highly polymorphic eukaryotic genome (Dehal et al. 2002; Vinson et al. 2005; Donmez and Brudno 2011; Kajitani et al. 2014; Safonova et al. 2015), assembly of *many* closely related bacterial strains can be substantially more difficult. While some studies described the initial steps toward identification and documenting of complex strain variants (Peng et al. 2011; Koren et al. 2011; Nijkamp et al. 2013), popular metagenomics assembly tools (Peng et al. 2012; Boisvert et al. 2012; Li et al. 2016) still include only rudimentary procedures for assembling strain-mixtures with high level of microdiversity.

We note that each of the challenges described above has already been addressed in the course of development of the SPAdes assembly toolkit, albeit in an application domain outside the field of metagenomics. SPAdes was initially developed to assemble datasets with non-uniform coverage, one of the key challenges of single cell assembly (Bankevich et al. 2012; Nurk et al. 2013). exSPAnder repeat resolution module of SPAdes (Prjibelski et al. 2014; Vasilinetc et al. 2015; Antipov et al. 2016) was developed to accurately resolve genomic repeats by combining multiple libraries sequenced with various technologies. Lastly, dipSPAdes (Safonova et al. 2015) was developed to address the challenge of assembling a two-haplome mixture within a highly-polymorphic diploid genome.

While these recently developed SPAdes tools address challenging assembly problems, metagenomics assembly is arguably an even more difficult problem with dataset sizes that dwarf most other DNA sequencing projects. Nevertheless, despite the fact that SPAdes was not designed for metagenomics applications, various groups successfully applied it to their metagenomics studies (Nurk et al. 2013; McLean et al. 2013; Coates et al. 2014; Kleigrewe et al. 2015; Bertin et al. 2015; Kleiner et al. 2015). While SPAdes indeed works well for assembling low complexity metagenomes like cyanobacterial filaments (Coates et al. 2014) or MDA-amplified mixtures of a small number of randomly selected bacterial cells (Nurk et al. 2013), its performance deteriorates in the case of complex bacterial communities.

Our novel metaSPAdes software combines new algorithmic ideas with proven solutions from the SPAdes toolkit to address various challenges of metagenomics assembly. Below we describe algorithmic approaches used in metaSPAdes and benchmark it against state-of-the-art metagenomics assemblers.

## Results

**Outline of metaSPAdes pipeline.** metaSPAdes first constructs the *de Bruijn graph* of all reads using SPAdes, transforms it into the *assembly graph* using various graph simplification procedures, and reconstructs paths in the assembly graph that correspond to long fragments of individual genomes within a metagenome (Bankevich et al. 2012; Nurk et al. 2013). Responding to the microdiversity challenge, metaSPAdes focuses on reconstructing a consensus backbone of a strain-mixture and thus sometimes ignores some strain-specific features (often corresponding to rare strains) to improve the contiguity of assemblies.

**Benchmarking challenges**. While genome assemblers are usually benchmarked on isolates with known reference genomes using various metrics (Salzberg et al. 2012; Gurevich et al. 2013), benchmarking of metagenomics assemblers is a more difficult task because no *reference metagenomes* are available for microbial communities of even moderate complexity. Moreover, even if the reference metagenome was known, benchmarking metagenomics assemblers would be a non-trivial task because metagenomic assemblers often discard information about rare species and strain variants to improve the contiguity of assembly. To address this trade-off, the quality assessment criteria should depend on the needs of a particular application. Unfortunately, until the community agrees on some standards for analyzing rare species and documenting strain variants, proper comparison of metagenomics assemblers will remain difficult.

Unavailability of the reference metagenomes is typically addressed by generating *synthetic* datasets with known community members. Such datasets can be obtained by sequencing the mixtures of bacteria with known genomes (Shakya et al. 2013; Turnbaugh et al. 2007), mixed from isolate sequencing data (Mavromatis et al. 2007) or simulated from reference sequences (Richter et al. 2008; Mende et al. 2012). However, while synthetic datasets proved to be useful in benchmarking various assemblers, they are typically less complex than real metagenomes (Koren et al. 2011; Peng et al. 2012).

Another approach to benchmarking metagenomics assemblers is based on identifying reference genomes closely related to some genomes in a metagenome (Treangen et al. 2013). However, this approach is limited since (i) related reference genomes are available only for a fraction of species in a metagenome, and (ii) differences between genomes in a metagenome and related (but not identical) references are often misinterpreted as assembly errors.

Recently, Mikheenko et al. (2016) developed metaQUAST, the first metagenomics-specific tool for evaluating assemblies. We used metaQUAST (from QUAST v4.0 package) to benchmark metaSPAdes (from SPAdes v3.8.2 package) against three popular metagenomics assemblers IDBA-UD v1.1.1 (Peng et al. 2012), Ray-Meta v2.3.1 (Boisvert et al. 2012) and MEGAHIT v1.0.5 (Li et al. 2015) across four diverse datasets.

**Datasets.** We analyzed the following datasets:

*Synthetic community dataset (SYNTH).* SYNTH is a set of reads from the genomic DNA mixture of 64 diverse bacterial and archaeal species (Shakya et al. (2013); SRA acc. no. SRX200676) that was used for benchmarking the Omega assembler (Haider et al. 2014). It contains 109 million Illumina HiSeq 100 bp paired-end reads with mean insert size of 206 bp. Since the reference genomes for all 64 species forming the SYNTH dataset are known, we used them to assess the quality of various SYNTH assemblies.

*CAMI simulated dataset (CAMI).* "Critical Assessment of Metagenome Interpretation" (CAMI) is a community initiative aimed at evaluating various approaches for analyzing metagenomes (http://www.cami-challenge.org/). Within this initiative, multiple synthetic datasets were simulated from reference genomes (including groups of closely related strains) to facilitate benchmarking of metagenomics pipelines. We used one of the "medium complexity" datasets simulated from 225 genomes (referred to as CAMI) and containing 150 million 100 bp paired-end reads with mean insert size of 180 bp (the errors in simulated reads are modelled after Illumina HiSeq reads).

*Human Microbiome Project dataset (HMP).* HMP is a female tongue dorsum dataset (SRA acc. no. SRX024329) generated by the the Human Microbiome Project (Huttenhower et al. 2012) that was used for benchmarking in Peng et al. (2011); Treangen et al. (2013); Mikheenko et al. (2016). It contains 75 million Illumina HiSeq 95 bp paired-end reads with mean insert size of 213 bp. Although the genomes comprising the HMP sample are unknown, we cautiously selected three reference genomes that are similar to the genomes within the sample for benchmarking.

*Soil metagenome dataset (SOIL).* Sharon et al. (2015) used both the TruSeq Synthetic Long Reads (TSLRs) (Kuleshov et al. 2014; McCoy et al. 2014) and conventional short reads to analyze complex soil metagenomic samples collected in an aquifer adjacent to the Colorado River. Since the TSLR technology generates unusually long metagenomics contigs (Kuleshov et al. 2015; Bankevich and Pevzner 2016; Dupont et al. 2016), these experiments provide a unique opportunity to benchmark various metagenomic assemblers based on how well they reconstruct genomic regions captured by the long TSLR contigs. We analyzed the dataset collected at depth of 4 meters (referred to as SOIL dataset) that contains 32 million Illumina HiSeq 150 bp paired-end reads with mean insert size of 460 bp. We further compared assemblies of the SOIL dataset against the set of scaffolds, resulting from TSLR reads assembled by truSPAdes in Bankevich and Pevzner (2016).

**Assembly parameters.** IDBA-UD was launched with read error-correction enabled as recommended in the manual for metagenomic data. Ray-Meta was launched with *k*-mer size equal to 31. All assemblers have been launched in 16 threads with default parameters. Supplementary Table S1 provides information about the running time and memory footprints of various assemblers.

| dataset/ assembler | metaSPAdes | | | MEGAHIT | | | IDBA-UD | | | Ray-Meta | | |
|---|---|---|---|---|---|---|---|---|---|---|---|---|
| | 10 | 1000 | ALL | 10 | 1000 | ALL | 10 | 1000 | ALL | 10 | 1000 | ALL |
| SYNTH | **9.6** | **120** | **196.7** | 6.1 | 103.7 | 196 | 6.9 | 111.7 | **196.7** | 5.8 | 93.1 | 183.3 |
| CAMI | **8** | **99.6** | 321.3 | 5.4 | 91.7 | 329.1 | 6.8 | 98.1 | **332** | 6.6 | 77.2 | 182.9 |
| HMP | **3.9** | **35.4** | 73.4 | 3 | 26.5 | 74.4 | 3.4 | 29 | **77.3** | 2.4 | 33.4 | 68.2 |
| SOIL | **1** | 19.3 | **203.7** | 0.4 | 10.5 | 145.6 | 0.9 | **19.9** | 168.6 | 0.3 | 4.1 | 11.1 |

Table 1. The total length of scaffolds generated by metaSPAdes, MEGAHIT, IDBA-UD, and Ray-Meta (in megabases) for SYNTH, CAMI, HMP, and SOIL datasets. Statistics are shown for 10 longest, 1000 longest, and all scaffolds longer than 1 kb. The best results for every dataset among all assemblers are highlighted in bold.

| dataset/assembler | metaSPAdes | MEGAHIT | IDBA-UD | Ray-Meta |
|---|---|---|---|---|
| SYNTH | **89.6 (125.7)** | 87.6 (122.2) | 88.5 (123.6) | 78.2 (109.1) |
| CAMI | 131.7 (176.6) | 134.4 (180) | **136.1 (182.7)** | 74.5 (100.2) |
| HMP | **28.4 (38.9)** | 26.5 (34.8) | 27.8 (36.8) | 27.5 (37.5) |
| SOIL | **61.5 (74.7)** | 41.1 (48.2) | 52.3 (63.9) | 3.3 (4) |

Table 2. Number (in thousands) and total length (in Mb) of predicted genes longer than 800 nt for all datasets and all assemblers.

| dataset/ assembler | | metaSPAdes | MEGAHIT | IDBA-UD | Ray-Meta |
|---|---|---|---|---|---|
| SYNTH | Fraction of aligned single reads | 96,38 | 93,60 | 96,17 | 93,98 |
| | Fraction of aligned paired reads (unique) | 91,67 | 88,36 | 91,71 | 84,76 |
| | Fraction of aligned paired reads (non-unique) | 2,94 | 3,27 | 2,58 | 7,62 |
| CAMI | Fraction of aligned single reads | 94,42 | 94,41 | 95,02 | 85,53 |
| | Fraction of aligned paired reads (unique) | 92,97 | 91,62 | 92,73 | 82,63 |
| | Fraction of aligned paired reads (non-unique) | 1,06 | 2,37 | 1,96 | 2,54 |
| HMP | Fraction of aligned single reads | 55,61 | 44,62 | 48,98 | 57,51 |
| | Fraction of aligned paired reads (unique) | 48,48 | 30,23 | 35,80 | 34,14 |
| | Fraction of aligned paired reads (non-unique) | 4,94 | 11,25 | 9,06 | 22,08 |
| SOIL | Fraction of aligned single reads | 16,80 | 13,04 | 15,37 | 2,67 |
| | Fraction of aligned paired reads (unique) | 13,65 | 10,06 | 12,14 | 1,77 |
| | Fraction of aligned paired reads (non-unique) | 0,08 | 0,11 | 0,23 | 0,05 |

Table 3. Fraction of aligned single and paired reads (both unique and non-unique) for all datasets and all assemblers (in percents). The colors of the cells reflect how much the results of various assemblers differ from the median value (blue/red cells indicate that the results improve/deteriorate as compared to the median value).

**Benchmarking.** Table 1 and Figure 1 provide the scaffold statistics and the cumulative scaffold length plots for all analyzed datasets. Ray-Meta generated inferior results compared to other assemblers on all datasets. metaSPAdes significantly improves the assembly in the case of the most complex SOIL dataset (26% and 47% increase in the total length of scaffolds as compared to IDBA-UD and MEGAHIT, respectively). To further evaluate how fragmented the resulting assemblies are, we used MetaProdigal v2.6.2 (Hyatt et al. 2012) to predict the complete genes in each assembly. Table 2 reports the number predicted genes longer than 800 bp (the gene length threshold is set to filter less reliable short gene predictions). Table 2 illustrates that, in the case of the most complex SOIL dataset, the number of predicted long genes in the metaSPAdes assembly is significantly larger as compared to other assemblers (18% and 49% increase as compared to IDBA-UD and MEGAHIT, respectively). Supplementary Text: "The summary of Nx statistics" presents Nx plots across all datasets (see metaQUAST manual for details on the Nx statistics).

For each dataset and assembler, we further aligned all single reads to all scaffolds (longer than 1kb) with bowtie2 (Langmead and Salzberg 2012). A read-pair is classified as *aligned* if both reads align to a single scaffold within 1kb from each other (with proper orientation). We further distinguished between uniquely and non-uniquely aligned read-pairs and reported the fractions of aligned single reads and read-pairs for all datasets and assemblers. Table 3 illustrates that 83%, 87%, 85%, 97% of reads in the SOIL dataset do not align to assemblies generated by SPAdes, MEGAHIT, IDBA-UD and Ray-Meta, respectively. Note that metaSPAdes assemblies have lower rates of non-unique paired-read alignments, indicating that they are less redundant.

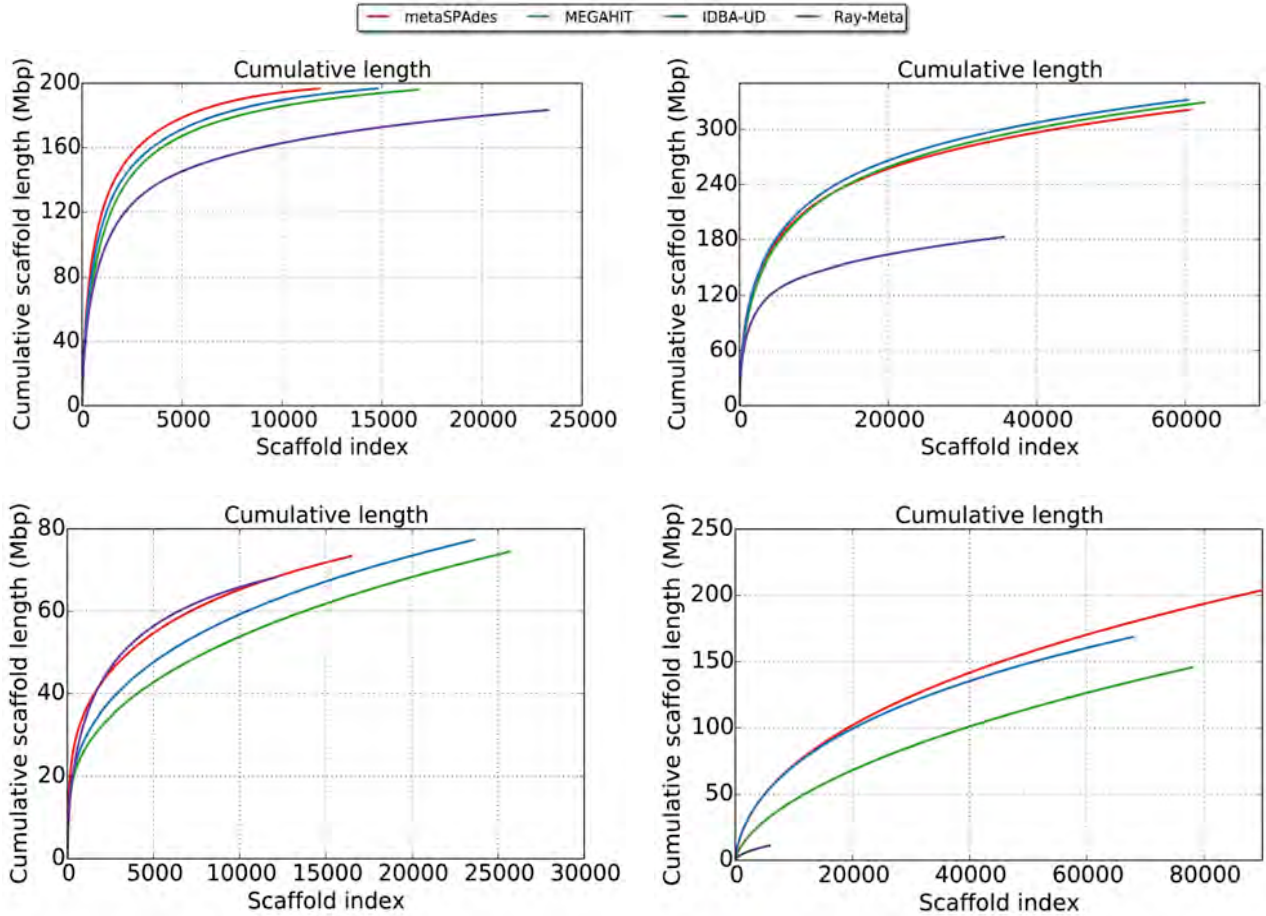

Figure 1. The cumulative scaffold lengths for SYNTH (top left), CAMI (top right), HMP (bottom left), and SOIL (bottom right) datasets. On the x-axis, scaffolds are ordered from the longest to the shortest. The y-axis shows the total length of x longest scaffolds in the assembly.

Below we discuss benchmarking results for each dataset in more details. Using metaQUAST output, we classify a position in a scaffold as an *intragenomic misassembly* if its flanking regions are aligned to non-consecutive regions of the same reference genome, and as *intergenomic misassembly* if they are aligned to different reference genomes or one of them remained unaligned. We also report the NGA50 statistics (NG50 corrected for assembly errors) to evaluate the quality of assembly of individual genomes within a metagenome. To compute NGA50 the contigs are first broken into smaller segments at the identified misassembly breakpoints. NGA50 for a given reference genome is the maximal value such that the broken segments (that align to this reference) of at least that length cover at least half of the bases of the reference.

*SYNTH dataset.* Figure 2 shows the results of benchmarking of various assemblers with respect to 20 most abundant species in the SYNTH dataset (see Table S2 for details) and reveals significant differences in performance across various assemblers even for this rather simple dataset. See Supplementary Text "Analysis of the SYNTH dataset" for the results on all references in the SYNTH dataset.

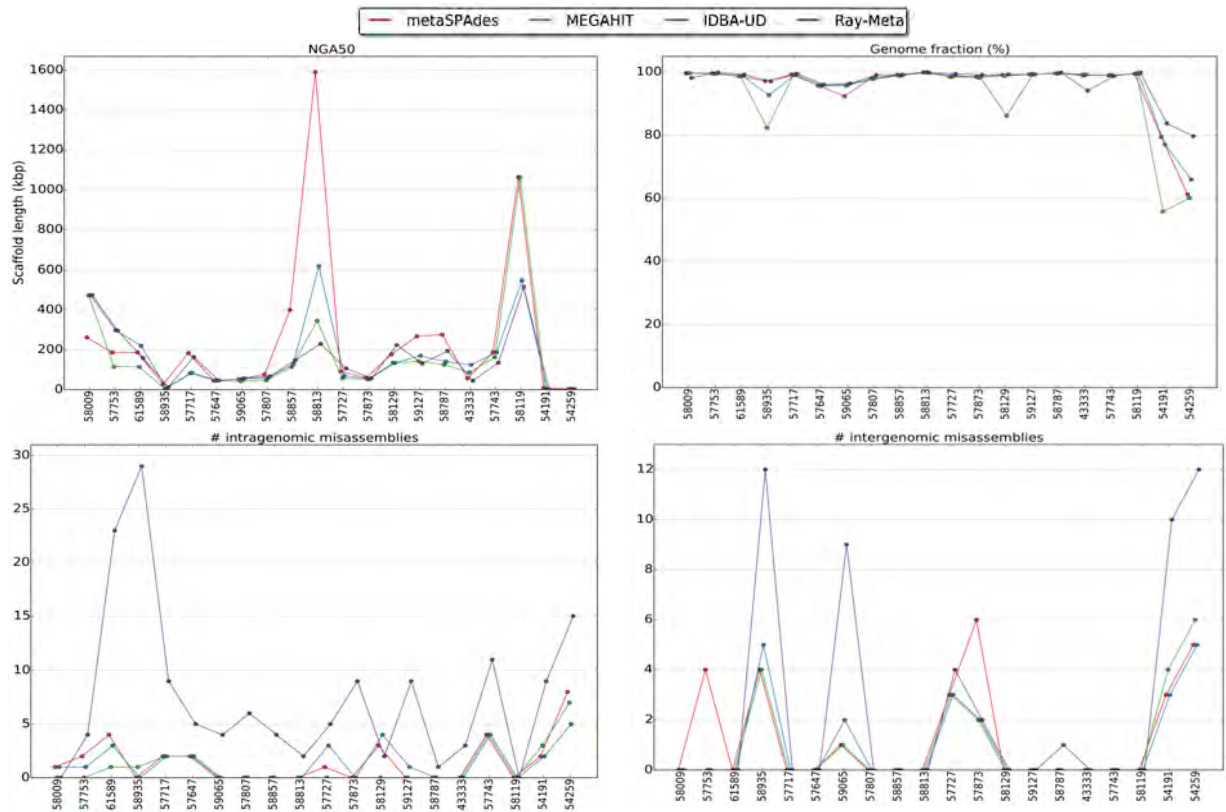

Figure 2. The NGA50 statistics (top left), the fraction of the reconstructed genome as compared to the total genome length (top right), the number of intragenomic misassemblies (bottom left) and the number of intergenomic misassemblies (bottom right) for 20 most abundant species comprising the SYNTH dataset. References are denoted by their RefSeq IDs (see Table S2) and arranged in the decreasing order of the coverage depths.

*CAMI dataset*. We analyzed the CAMI dataset with respect to 20 most abundant reference genomes for this dataset. See Table S4 for the list of these species and Supplementary Text "Analysis of the CAMI dataset" for the assembly statistics for each of them.

*HMP dataset*. Since the genomes comprising bacterial communities are typically unknown, HMP consortium identified a number of reference genomes (listed at HMP Shotgun Community profiling SRS077736) similar to the genomes comprising the HMP dataset (Treangen et al. 2013). However, our attempt to use this resource for reliable quality assessment faced difficulties: only three genomes in this list (*Streptococcus salivarius SK126, Neisseria subflava NJ9703*, and *Prevotella melaninogenica ATCC 25845*) were at least 70% covered by contigs generated by assemblers included in this study. Moreover, we revealed substantial differences between these references and the genomes in the sample, making metaQUAST analysis unreliable (see Supplementary Text "Analysis of the HMP dataset").

*SOIL dataset.* As discussed in Sharon et al. (2015), due to the complexity of the SOIL dataset, assemblies of short-reads and TSLR data are not expected to have a large overlap. Indeed, since short read assemblies deteriorate with the decrease in the genome coverage,, contigs from rare genomes are typically under-represented in metagenomics assemblies. TSLRs, on the other hand, are expected to capture large fragments from rare genomes within a metagenome (Bankevich and Pevzner 2016). Consistent with this observation, Sharon et al. (2015) found that majority of TSLRs originated from genomes with less than 5x coverage by short-read Illumina libraries.

We compared assemblies of the SOIL dataset against the set of contigs obtained from the TSLR data in Bankevich and Pevzner (2016) (which improves on the original TSLR assemblies in Sharon et al. (2015)). Contigs longer than 20 kb (total length 103 Mb) were selected as a "reference genome" for computing QUAST statistics (Tables 4 and 5). Only 27.6 Mb (≈13.6%) of the total length of the metaSPAdes scaffolds longer than 1 kb (196 Mb) overlapped with TSLR contigs, covering just ≈26% of the total length of the TSLR assembly.

|  | metaSPAdes | MEGAHIT | IDBA-UD | Ray-Meta |
|---|---|---|---|---|
| #misassemblies | 236 | 209 | 322 | 35 |
| Percentage of length of the TSLR contigs covered by the metagenomics contigs | 25.9 | 21.6 | 24.4 | 4.5 |
| Total length of the metagenomics assembly not aligned to the TSLR contigs (Mb) | 176.0 | 123.0 | 142.1 | 6.2 |

Table 4. Comparison of long contigs (longer than 1 kb) generated by various metagenomics assemblers for SOIL dataset against TSLR contigs generated in Bankevich and Pevzner (2016). The total length of long contigs generated by metaSPAdes significantly exceeds the total length of long contigs generated by other metagenomics assemblers.

| coverage/assembler | metaSPAdes | | | | MEGAHIT | | | | IDBA-UD | | | | Ray-Meta | | | |
|---|---|---|---|---|---|---|---|---|---|---|---|---|---|---|---|---|
| | total length | #contigs | #errors | #errors/Mb | total length | #contigs | #errors | #errors/Mb | total length | #contigs | #errors | #errors/Mb | total length | #contigs | #errors | #errors/Mb |
| < 5X | 89.9 | 53.6 | 24 | 0.3 | 43.6 | 30.3 | 16 | 0.4 | 51.1 | 31.3 | 29 | 0.6 | 0 | 0 | 0 | 0 |
| 5-10X | 83.1 | 28.8 | 83 | 1 | 80 | 36.6 | 76 | 1 | 87.3 | 28.9 | 125 | 1.4 | 1.5 | 1.1 | 0 | 0 |
| 10-15X | 17.2 | 4.5 | 57 | 3.3 | 14.5 | 6.6 | 52 | 3.6 | 18.4 | 4.8 | 88 | 4.8 | 2.6 | 1.4 | 6 | 2.3 |
| 15-20X | 9.5 | 1.8 | 45 | 4.7 | 6.9 | 3.1 | 40 | 5.8 | 8.7 | 1.9 | 49 | 5.6 | 4.7 | 2.3 | 13 | 2.8 |
| > 20X | 3.8 | 0.8 | 27 | 7.1 | 2.5 | 1.3 | 24 | 9.6 | 3 | 0.9 | 30 | 10 | 2.3 | 1 | 16 | 7 |

Table 5. Comparison of long contigs (longer than 1 kb) generated by various metagenomics assemblers for SOIL dataset. Contigs were divided into bins by their average coverage and compared against TSLR contigs constructed in Bankevich and Pevzner (2016). Total length (in Mb), the number of contigs (in thousands), the number of misassemblies and the number of misassemblies per Mb are shown for each bin and each assembler. The reason for the surprising increase in the number of misassemblies per Mb with increase in the coverage remains unclear. One possible explanation is that the average contigs length increases with the coverage and long contigs are more prone to misassemblies.

## Discussion

metaSPAdes addressed a number of challenges in metagenomics assembly and implemented several novel features (see Methods section), such as:

- an efficient approach to analyzing microdiversity,

- a new repeat resolution pipeline that, somewhat counter-intuitively, utilizes rare strain variants to improve the consensus assembly of strain-mixtures,

- fast algorithms for construction and simplification of the de Bruijn graphs as well as error-correction of reads.

These features contributed to improvements in metaSPAdes assemblies (as compared to the state-of-the-art assemblers MEGAHIT, IDBA-UD and Ray-Meta) and enabled us to scale metaSPAdes for analyzing large metagenomes.

In addition to the intrinsic *biological* challenges discussed in this paper, the field of metagenomics assembly also faces *technological* challenges caused by innovations in sequencing and library preparation techniques. For example, recently introduced high-quality jumping (mate-pair) libraries (such as *Nextera Mate Pair Libraries*) have a potential to significantly improve assembly quality (Vasilinetc et al. 2015). However, metagenomics assemblers have not caught up with this technological innovation yet. Another example is the TSLR technology (Kuleshov et al. 2014; McCoy et al. 2014), whose first metagenomics applications highlighted the need for developing methods to reliably combine it with the paired-end libraries (Sharon et al. 2015; Kuleshov et al. 2015; Bankevich and

Pevzner 2016). metaSPAdes now faces the challenge of incorporating these emerging technologies into its assembly pipeline.

## Methods

**Detecting and masking strain variations**. Genomic differences between related strains often result in *bulges* and *tips* in the de Bruijn graphs that are not unlike artifacts caused by sequencing errors in traditional genome assembly (Pevzner et al. 2004; Zerbino and Birney 2008). For example, a sequencing error often results in a bulge formed by two alternative paths of similar lengths between the same vertices in the de Bruijn graph, a "correct" path with high coverage and an "erroneous" path with low coverage. Similarly, a substitution or a small indel in a rare strain (as compared to an abundant strain) often results in a bulge formed by a high-coverage path corresponding to the abundant strain and an alternative low-coverage path corresponding to the rare strain.

Aiming at the consensus assembly of a strain-mixture, metaSPAdes masks the majority of strain differences using a modification of the SPAdes procedures for masking sequencing errors (the algorithms for removal of tips, *simple bulges* (Bankevich et al. 2012), and *complex bulges* (Nurk et al. 2013)). metaSPAdes uses more aggressive settings than the ones used in assemblies of isolates, e.g. it collapses larger bulges and removes longer tips than SPAdes. We note that the bulge *projection* approach in SPAdes improves on the originally proposed bulge *removal* approach (Pevzner et al. 2004; Zerbino and Birney 2008) used in most existing assemblers since it stores valuable information about the processed bulges. This feature is important for the repeat resolution approach in metaSPAdes described below.

**Analyzing filigree edges in the assembly graph.** In addition to single nucleotide variations and small indels, strain variations are often manifested as highly diverged regions, insertions of mobile elements, rearrangements, large deletions, parallel gene transfer, etc. The green edges in the assembly graph shown in Figure 4 result from an additional copy of a mobile element in a rare *strain$_2$* (compared to the abundant *strain$_1$*) while the blue edge corresponds to a horizontally transferred gene (or a highly diverged genomic region) in a rare *strain$_3$* (compared to the abundant *strain$_1$*). Such edges fragment contigs corresponding to the abundant *strain$_1$*, e.g., the green edges in Figure 4 (bottom right) break the edge *c* into three shorter edges. We note that the edges in the assembly graph are *condensed*, i.e., they represent non-branching paths formed by *k*-mers.

It is not immediately clear how to analyze the low coverage edges resulting from such large-scale strain variations within the strain mixture. We refer to edges originating from rare strain variants within the assembly graph of a strain-mixture as *filigree edges*. Traditional genome assemblers use a *global* threshold on read coverage to remove the low coverage edges (that typically result from sequencing errors) from the assembly graph during the graph simplification step. However, this approach does not work well for metagenomics assemblies, since there is no global threshold that (i) removes edges corresponding to rare *strains* and (ii) preserves edges corresponding to rare *species*. Similarly to IDBA-UD and MEGAHIT, metaSPAdes analyzes the *coverage ratios* between adjacent edges in the assembly graph, classifying edges with low coverage ratios as potential filigree edges.

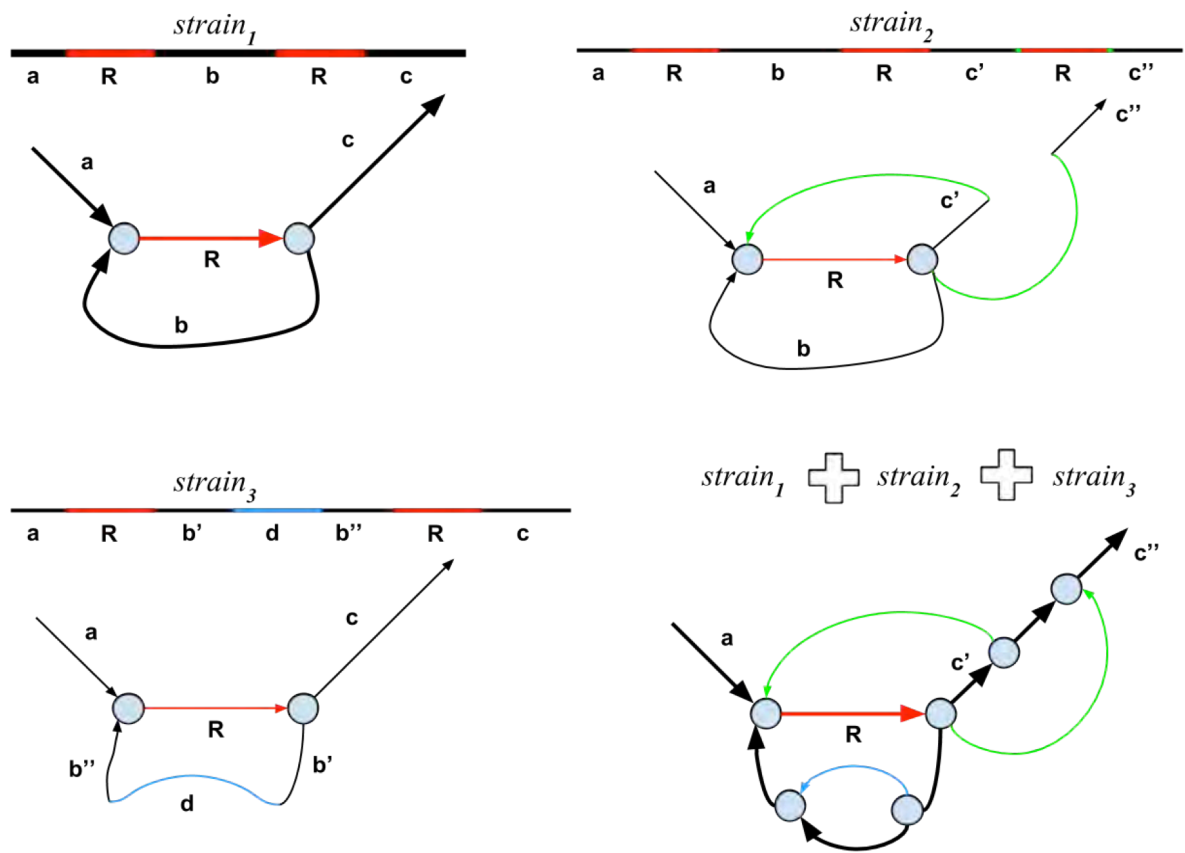

Figure 4. The de Bruijn graphs of three strains and their strain mixture. The abundant strain ($strain_1$) is shown by thick lines and the rare strains ($strain_2$ and $strain_3$) are shown by thin lines. The genomic repeat R is shown in red. (Upper Left) The de Bruijn graph of the abundant $strain_1$. (Upper Right) The rare $strain_2$ differs from the abundant $strain_1$ by an insertion of an additional copy or repeat R. The two breakpoint edges resulting from this insertion are shown in green. These filigree edges are not removed by the graph simplification procedures in the standard assembly tools aimed at isolates. (Bottom Left) The rare $strain_3$ differs from the abundant $strain_1$ by an insertion of a horizontally transferred gene (or a highly diverged genomic region). (Bottom Right) The de Bruijn graph of the mixture of three strains.

We denote the coverage of an edge *e* in the assembly graph as *cov(e)* and define the coverage *cov(v)* of a vertex *v* as the maximum of *cov(e)* over all edges *e* incident to *v*. Given an edge *e* incident to a vertex *v* and a threshold *ratio* (the default value is 10), a vertex *v* *predominates* an edge *e* if its coverage is significantly higher than the coverage of the edge *e*, i.e., if $ratio \cdot cov(e) < cov(v)$. An edge (*v,w*) is *weak* if it is predominated by either *v* or *w*. Note that filigree edges are often classified as weak since their coverage is much lower than the coverage of adjacent edges resulting from abundant strains.

metaSPAdes *disconnects* all weak edges from their predominating vertices in the assembly graph. Disconnection of a weak edge (*v,w*) in the assembly graph from its starting vertex *v* (ending vertex *w*) is simply a removal of its first (last) *k*-mer rather than removal of the entire edge. We emphasize that, in difference from IDBA-UD and MEGAHIT, we *disconnect* rather than *remove* weak edges in the assembly graph since our goal is to preserve the information about rare strains whenever possible, i.e., when it does not lead to a deterioration of the consensus backbone.

**Repeat resolution with exSPAnder.** exSPAnder (Prjibelski et al. 2014; Vasilinetc et al. 2015; Antipov et al. 2016) is a module of SPAdes that combines various sources of information (e.g., paired reads or long error-prone reads) for resolving repeats and scaffolding in the assembly graph. Starting from a path consisting of a single condensed edge in the assembly graph, exSPAnder iteratively attempts to extend it into a longer *genomic path* that represents a contiguous segment of the genome. To extend a path, exSPAnder selects one of its *extension edges* (all the edges that start at the terminal vertex of this path). Choice of the extension edge is controlled by the *decision rule* that evaluates whether a particular extension edge is sufficiently supported by the data, while other extension edges are not (given the existing path). exSPAnder further removes overlaps (*overlap reduction* step of exSPAnder) between generated genomic paths and outputs the strings spelled by the resulting paths as a set of contigs.

metaSPAdes modifies the decision rule of exSPAnder to account for the *local* read coverage, denoted *localCov*, of the *specific* genomic region that is being reconstructed during the path extension process. See Supplementary Text "Modifying the decision rule in exSPAnder for metagenomics data" for details. The value *localCov* is estimated as the minimum across the average coverages of the edges in the path that is being extended. Taking minimum (rather than the average) coverage excludes the repetitive edges in the path from consideration and typically underestimates the real coverage of the region, making the decision rule more conservative.

**A new metagenomics decision rule in metaSPAdes.** Some intergenomic repeats between species of different abundances can be resolved based on the differences in the depth of read coverage (Namiki et al. 2012; Haider et al. 2014). metaSPAdes introduces an additional metagenomics-specific decision rule that filters out unlikely path extensions using the coverage estimate of the region that is being reconstructed (Figure 5). It often allows metaSPAdes to pass through long inter-species repeats during reconstruction of abundant species. metaSPAdes applies a new decision rule described below only if the paired reads failed to provide sufficient evidence to discriminate between extension edges.

An edge in the assembly graph is called *long* if its length exceeds a certain threshold (1500 bp by default) and *short* otherwise. We say that a long edge $e_2$ *follows* a long edge $e_1$ in a genomic path if all edges between the end of $e_1$ and the start of $e_2$ in this path are short.

While considering an extension edge $e$, metaSPAdes performs a directed traversal of the graph (Figure 5b), starting from the end of $e$ and walking along the short edges. We define the set of all vertices that are reached by this traversal as *frontier(e)* and consider the set *next(e)* of all long edges starting in *frontier(e)*. This procedure is aimed at finding a non-repetitive long edges that can follow $e$ in the (unknown) genomic path. We classify an edge in the set *next(e)* as a *low-coverage* edge if the coverage estimate of the region that is being reconstructed, *localCov,* exceeds its coverage at least by a factor $\beta$ (the default value $\beta$=2). If all edges in *next(e)* are low-coverage edges, then $e$ is considered an unlikely candidate for an extension of the current path. If all but a single edge $e'$ represent unlikely extensions, the path is extended by $e'$ (Figure 5c).

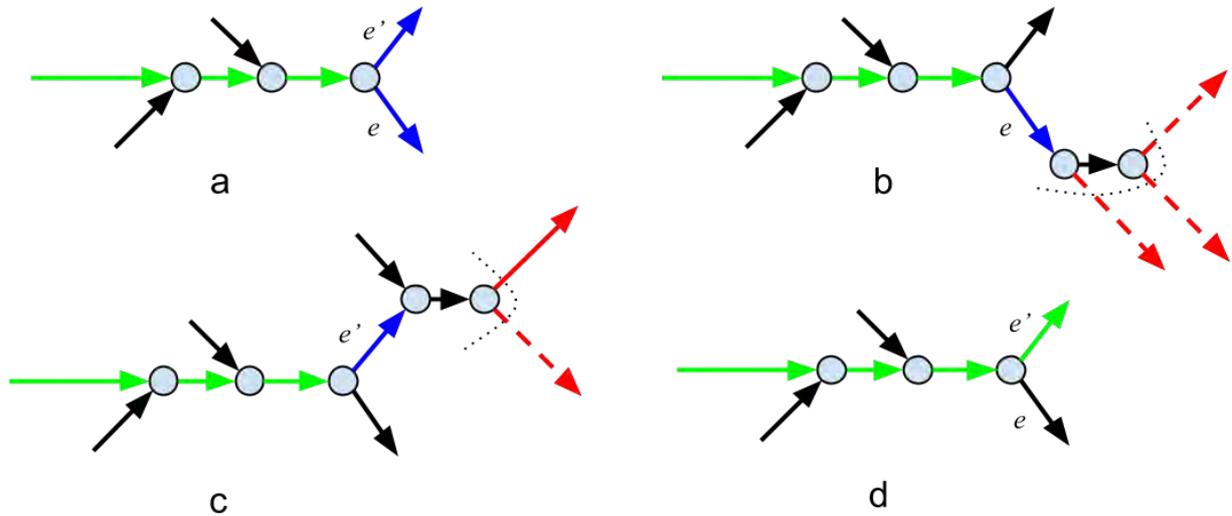

Figure 5. Applying the metagenomics-specific decision rule for repeat resolution. a) The path that is currently being extended (formed by green edges) along with its blue extension edges *e and e'*. b) The short-edge traversal from the end of the extension edge *e*. The dotted curve shows the boundary *frontier(e)* of the traversal. The edges in the set *next(e)* are shown in red with low-coverage edges represented as dashed arrows (other edges in *next(e)* are represented as solid arrows). Since all edges in *next(e)* have low coverage, the edge *e* is ruled out as an unlikely extension candidate. c) The short-edge traversal from the end of the extension edge *e'*. d) Since *e'* is a single extension edge that was not ruled out (there is a solid edge in *next(e')*), it is added to the growing path and the extension process continues.

**Utilizing strain differences for repeat resolution in metaSPAdes.** Safonova et al. (2015) showed that differences between haplomes can be used to improve the quality of consensus assembly of a highly polymorphic diploid genome. metaSPAdes capitalizes on the similar observation that the differences between strains can be used to improve the quality of consensus assembly of a strain-mixture. In particular, *strain-contigs* (generated prior to masking strain-differences in assembly graph) representing individual strains often provide additional long-range information for reconstruction of a strain-mixture backbone.

Inspired by dipSPAdes (Safonova et al. 2015), metaSPAdes uses the following pipeline that includes two launches of the exSPAnder module (Figure 6).

- *Generating strain-contigs.* After constructing the assembly graph (that encodes both abundant and rare strains), we launch exSPAnder to generate a set of strain-contigs representing both rare and abundant strains (Figure 6c). Strain-contigs are not subjected to the default overlap reduction step in exSPAnder.

- *Transforming assembly graph into consensus assembly graph.* metaSPAdes identifies and masks rare strain variants, resulting in the *consensus assembly graph* (Figure 6d).

- *Generating strain-paths in the consensus assembly graph.* Capitalizing on the bulge projection approach (Bankevich et al. 2012; Nurk et al. 2013), metaSPAdes reconstructs paths in the consensus assembly graph corresponding to strain-contigs, referred to as *strain-paths* (Figure 6e).

- *Repeat resolution using strain-paths.* This step utilizes the hybrid mode of exSPAnder originally developed to incorporate long error-prone Pacific Biosciences and Oxford Nanopore reads in the repeat resolution process (Antipov et al. 2016; Ashton et al. 2014; Labonté et al. 2015). Instead of working with long error-prone reads, we modified exSPAnder to work with virtual reads spelled by the strain-paths to facilitate resolution of repeats in the consensus assembly graph (Figure 6f).

Note that in the example in Figure 6, the long red repeat with multiplicity 2 in the abundant strain is resolved because of the variations (diverged green copy of the repeat) in the rare strain.

**Scaling metaSPAdes.** Since some metagenomics datasets contain billions of reads, metagenomics assemblers have to be optimized with respect to both speed and memory footprint (Nagarajan and Pop 2013). Supplementary Text "Reducing running time and memory footprint of metaSPAdes" describes efforts to scale metaSPAdes for assembling large metagenomic datasets.

## Acknowledgements

This work was supported by the Russian Science Foundation (grant 14-50-00069). We are grateful to Chris Dupont, Mihai Pop, and Bahar Behsaz for useful comments. We are also grateful to Alla Lapidus, who brought our attention to the field of metagenomics.

## Disclosure Declaration

Authors have no conflicts to report.

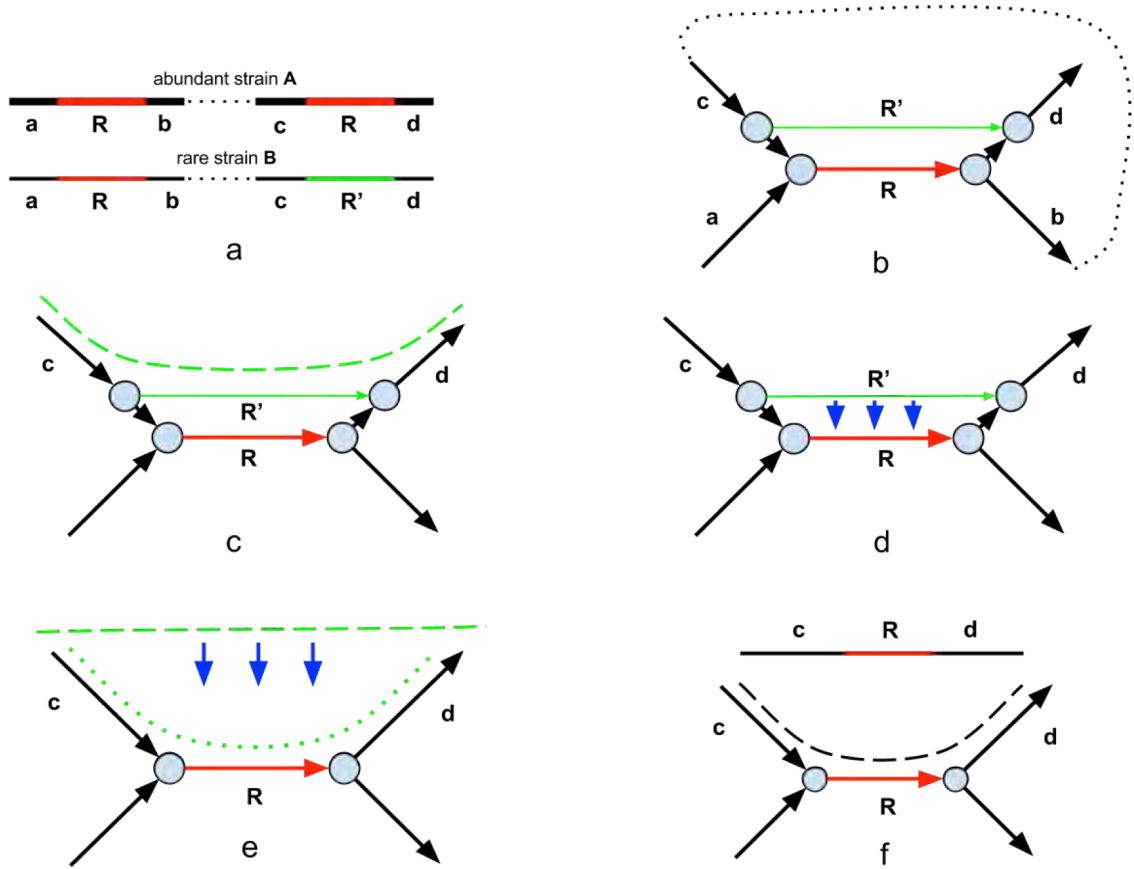

Figure 6. Repeat resolution in metagenomics assembly. a) One of two identical copies of a long (longer than the insert size) "red" repeat R in the abundant strain has mutated into a unique genomic "green" region R' in the rare strain. b) The assembly graph resulting from a mixture of reads from the abundant and rare strains. Two alternative paths between the start and the end of the green edge (one formed by a single green edge and another formed by two black and one red edge) form a bulge. c) The strain-contig spanning R' (shown by green dashed line) constructed by exSPAnder at the "Generating strain-contigs" step. d) Masking of the strain variations at the "Transforming assembly graph into consensus assembly graph" step leads to a projection of a bulge (formed by red and green edges) and results in the consensus assembly graph shown in the (e) panel. The blue arrows emphasize that SPAdes *projects* rather than *deletes* bulges, facilitating the subsequent reconstruction of strain-paths in the consensus assembly graph. (e) Reconstruction of the strain-path (green dotted line), corresponding to a strain-contig (green dashed line) at the "Generating strain-paths in the consensus assembly graph" step. f) At the "Repeat resolution using strain-paths" step, metaSPAdes utilizes both strain-paths and paired reads to resolve repeats in the consensus graph. The green dotted strain-path from the (e) panel is used as additional information to reconstruct the consensus contig *cRd* spanning the long repeat.

# metaSPAdes: a new versatile metagenomics assembler
## (Supplementary Material)

Sergey Nurk[1,*,**], Dmitry Meleshko[1,*], Anton Korobeynikov[1,2] and Pavel A Pevzner[1,3]

[1]Center for Algorithmic Biotechnology, Institute for Translational Biomedicine,
St. Petersburg State University, St. Petersburg, Russia

[2]Department of Statistical Modelling, St. Petersburg State University, St. Petersburg, Russia

[3]Department of Computer Science and Engineering, University of California,
San Diego, USA

*These authors contributed equally to this work
**corresponding author, sergeynurk@gmail.com

**Supplementary Text A: Running time and memory footprint of the benchmarked assemblers**

| dataset | metaSPAdes | MEGAHIT | IDBA-UD | Ray-Meta |
|---|---|---|---|---|
| SYNTH | 5h 28m (26.8) | 1h 20m (8.3) | 4h 37m (108.6) | 8h 17m (38.4) |
| CAMI | 17h 45m (130.9) | 2h 54m (11.2) | 8h 35m (557.6) | 15h 15m (68.9) |
| HMP | 4h 51m (21.7) | 1h 26m (7.3) | 4h 49m (234.5) | 5h 59m (26.9) |
| SOIL | 32h 57m (185.1) | 3h 17m (15.3) | 12h 49m (114.7) | 7h 79m (63.9) |

Table S1. The running time and memory footprint (in Gb) of various metagenomics assemblers.

**Supplementary Text B: Modifying the decision rule in exSPAnder for metagenomics data**

The decision rule in exSPAnder uses a binary *support function Support(e, e', D)* that reflects whether the read-pairs *connecting* edges *e* and *e'* support the conjecture that *e'* follows *e* at the distance *D* in the genome (see Prjibelski et al. (2014) and Vasilinetc et al. (2015) for details). exSPAnder automatically adjusts its support function to the particular dataset based on the average coverage for the *entire* dataset (in the case of isolate sequencing). However, since the support function is not adjusted to *local* coverage, exSPAnder is applying the same parameters to regions from both abundant and rare bacterial species, leading to suboptimal and error-prone metagenomics assemblies. metaSPAdes modifies the support function to take into account the read coverage *localCov* of the *specific* genomic region that is being reconstructed during the path extension process.

After *localCov* is computed (see section "Repeat resolution with exSPAnder" for details), metaSPAdes computes the following values based on the empirically estimated distribution of the insert sizes (see Prjibelski et al. (2014) and Vasilinetc et al. (2015) for details):

- *ExpectedReadPairs$_{localCov}$(e, e', D)*: the expected number of read-pairs connecting edges $e$ and $e'$ separated in the genome by distance $D$, under the assumption that the coverage is uniform with average value *localCov*. Given the distribution of insert sizes and *localCov*, the value *ExpectedReadPairs$_{localCov}$(e, e', D)* is defined by the lengths of edges $e$ and $e'$ and distance $D$.

- *ReadPairs(e,e',D)*: the total number of read-pairs from the metagenomics dataset that support the conjecture that $e'$ follows $e$ in the genome at distance $D$.

- *Support(e, e', D)* = 1 iff *ReadPairs(e,e',D)/ ExpectedReadPairs$_{localCov}$(e, e, D)* > $\alpha$ (the default value $\alpha=0.3$).

**Supplementary Text C: Reducing running time and memory footprint of metaSPAdes**

Most existing assembly tools fail when faced with analyzing large metagenomics datasets. For example, since all metagenomic assemblers available in 2014 failed to process a large soil dataset with 3.3 billion reads, Howe et al. (2014) preprocessed the data by performing digital normalization of coverage depth; constructing the probabilistic de Bruijn graph (Pell et al. 2012) based on the *Bloom filter* (Bloom 1970); and subdividing the set of reads based on partitioning of this graph. Chikhi and Rizk (2013) developed Minia assembler that efficiently represents the de Bruijn graphs using the Bloom filter (Chikhi and Rizk 2013; Salikhov et al. 2014). Li et al. (2015) used the concept of the *succinct de Bruijn graph* (Bowe et al. 2012) to develop a fast and memory-efficient MEGAHIT assembler.

metaSPAdes uses a different approach to address the speed and memory bottlenecks of metagenomics assemblies. Utilizing the state-of-the-art *perfect hashing* technique (Botelho et al. 2013), it implements a compact representation of the uncondensed de Bruijn graph as well as new efficient algorithms for its construction and simplification. Our use of perfect hashing for representing the de Bruijn graph differs from the previous approach in Chapman et al. (2011) that did not enable efficient de Bruijn graph simplification procedures. It also improves on the approach in Iqbal et al. (2012) by significantly reducing the memory footprint. We also parallelized the most time-consuming procedures for transforming the de Bruijn graph into the assembly graph (e.g., processing of bulges) and optimized the BayesHammer error-correction module of SPAdes (Nikolenko et al. 2013).

Since our approach to the de Bruijn graph representation and the abovementioned speed-ups apply to both SPAdes and metaSPAdes, they will be described elsewhere.

**Supplementary Text D: The summary of Nx statistics**

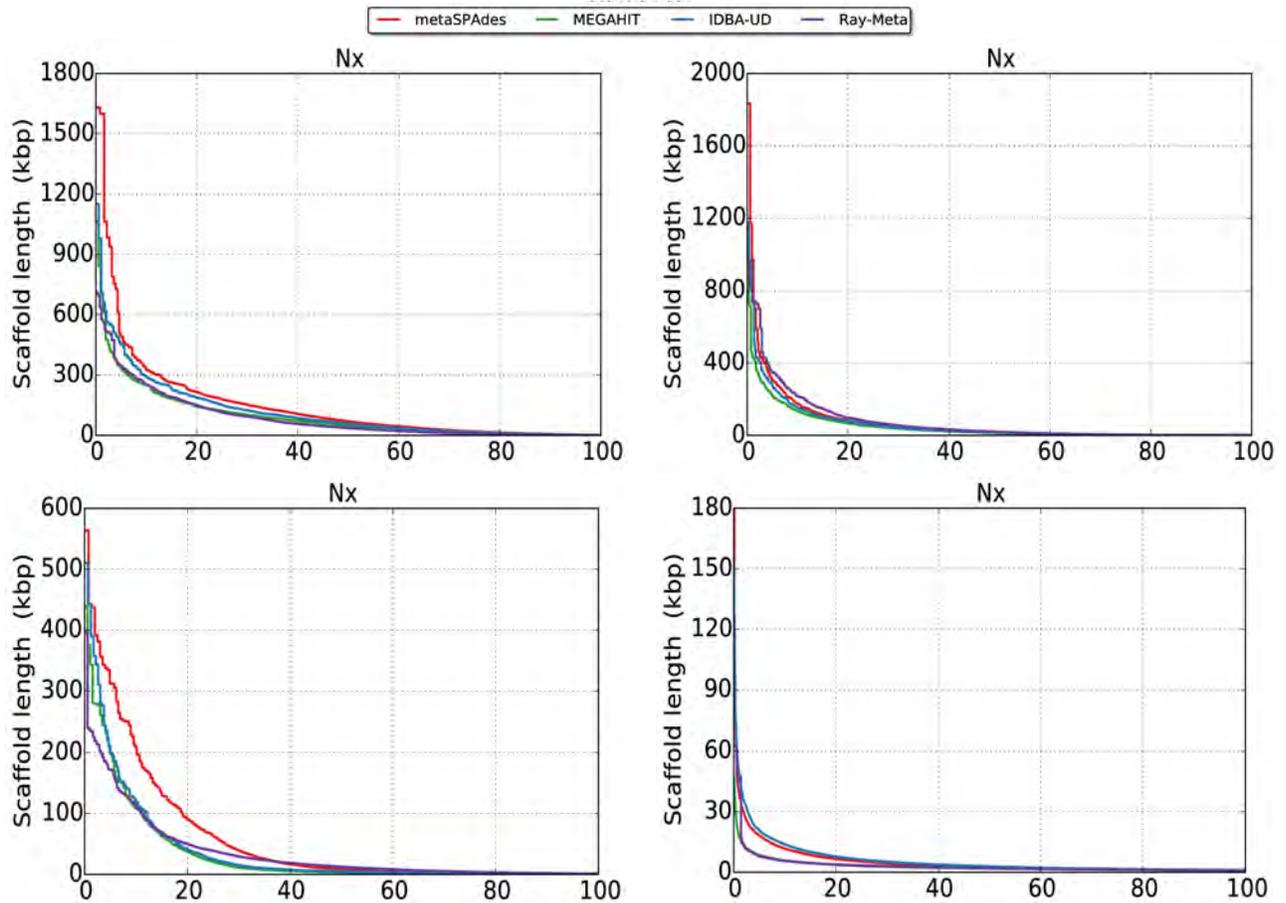

Figure S1. The Nx statistics for the SYNTH (top left), CAMI (top right), HMP (bottom left), and SOIL (bottom right) datasets. Nx is the length for which the collection of all scaffolds of that length or longer covers at least x percent of the total contig length in an assembly. For example, Nx for x=50 corresponds to the standard N50 value. Only scaffolds longer than 1 kb were considered for computing the Nx statistics.

## Supplemental Text E: Analysis of the SYNTH dataset

| No. | RefSeq ID | Species Name | Genome size (Mbp) | Average coverage | Abbreviation |
|---|---|---|---|---|---|
| 1 | 58009 | *Nanoarchaeum equitans* | 0,49 | 318 | Neq |
| 2 | 57753 | *Pyrococcus horikoshii* | 1,74 | 138 | Pho |
| 3 | 61589 | *Rhodopirellula baltica* | 7,15 | 137 | Rba |
| 4 | 58935 | *Thermotoga sp. RQ2* | 1,88 | 128 | ThRQ2 |
| 5 | 57717 | *Archaeoglobus fulgidus* | 2,18 | 124 | Afu |
| 6 | 57647 | *Nitrosomonas europaea* | 2,81 | 117 | Neu |
| 7 | 59065 | *Thermotoga neapolitana DSM 4359* | 1,88 | 112 | ThDSM4359 |
| 8 | 57807 | *Sulfolobus tokodaii* | 2,7 | 102 | Sto |
| 9 | 58857 | *Hydrogenobaculum sp. Y04AAS1* | 1,56 | 94 | HY04AAS1 |
| 10 | 58813 | *Gemmatimonas aurantiaca* | 4,64 | 90 | Gau |
| 11 | 57727 | *Pyrobaculum aerophilum IM2* | 2,22 | 90 | PaeIM2 |
| 12 | 57873 | *Pyrococcus furiosus* | 1,9 | 86 | Pfu |
| 13 | 58129 | *Chlorobium phaeovibrioides* | 1,97 | 84 | Cvi |
| 14 | 59127 | *Acidobacterium capsulatum* | 4,13 | 82 | Aca |
| 15 | 58787 | *Pyrobaculum calidifontis* | 2 | 80 | Pca |
| 16 | 43333 | *Aciduliprofundum boonei* | 1,49 | 78 | Abo |
| 17 | 57743 | *Geobacter sulfurreducens PCA* | 3,81 | 76 | GsuPCA |
| 18 | 58119 | *Persephonella marina EX-H1* | 1,98 | 74 | PmaEX-H1 |
| 19 | 54191 | *Sulfitobacter sp. EE-36* | 3,6 | 73 | SEE-36 |
| 20 | 54259 | *Sulfitobacter sp. NAS-14.1* | 4,03 | 72 | SNAS-14.1 |
| 21 | 57713 | *Methanocaldococcus jannaschii* | 1,74 | 65 | Mja |
| 22 | 57583 | *Treponema denticola* | 2,84 | 63 | Tde |
| 23 | 57883 | *Methanopyrus kandleri* | 1,69 | 62 | Mka |
| 24 | 58409 | *Pyrobaculum arsenaticum* | 2,12 | 55 | Pas |
| 25 | 54637 | *Sulfurihydrogenibium yellowstonense SS-5* | 1,53 | 55 | SyeSS-5 |
| 26 | 57897 | *Chlorobium tepidum* | 2,15 | 53 | Cte |
| 27 | 58741 | *Methanococcus maripaludis C5* | 1,81 | 51 | MmaC5 |
| 28 | 59177 | *Dictyoglomus turgidum* | 1,86 | 50 | Dtu |
| 29 | 58655 | *Thermotoga petrophila RKU-1* | 1,82 | 48 | TpeRKU-1 |
| 30 | 58223 | *Thermus thermophilus HB8* | 1,85 | 47 | TthHB8 |
| 31 | 58127 | *Chlorobium limicola* | 2,76 | 47 | Cli |
| 32 | 54519 | *Desulfovibrio piger* | 2,9 | 46 | DesPig |
| 33 | 58289 | *Caldicellulosiruptor saccharolyticus* | 2,97 | 44 | Csa |
| 34 | 61591 | *Wolinella succinogenes* | 2,11 | 44 | Wsu |
| 35 | 58035 | *Methanococcus maripaludis S2* | 1,66 | 43 | MmaS2 |
| 36 | 58133 | *Chlorobium phaeobacteroides* | 3,13 | 41 | Cph |
| 37 | 58173 | *Pelodictyon phaeoclathratiforme* | 3,02 | 38 | Pph |
| 38 | 57657 | *Chloroflexus aurantiacus J-10-fl* | 5,26 | 37 | CauJ-10-fl |
| 39 | 58985 | *Akkermansia muciniphila* | 2,66 | 35 | Amu |
| 40 | 57917 | *Clostridium thermocellum* | 3,84 | 34 | Cth |
| 41 | 58879 | *Porphyromonas gingivalis* | 2,35 | 33 | Pgi |
| 42 | 57669 | *Enterococcus faecalis* | 3,34 | 33 | Efa |
| 43 | 59201 | *Caldicellulosiruptor bescii* | 2,91 | 32 | Cbe |
| 44 | 58339 | *Thermoanaerobacter pseudethanolicus* | 2,36 | 28 | Tps |
| 45 | 58971 | *Leptothrix cholodnii* | 4,91 | 26 | Lch |
| 46 | 57803 | *Nostoc sp. PCC 7120* | 7,2 | 26 | NPCC7120 |
| 47 | 58679 | *Desulfovibrio vulgaris DP4* | 3,66 | 26 | DvuDP4 |
| 48 | 58365 | *Ignicoccus hospitalis* | 1,3 | 25 | Iho |
| 49 | 399 | *Bacteroides thetaiotaomicron* | 6,29 | 24 | Bth |
| 50 | 46845 | *Haloferax volcanii* | 2,85 | 24 | Hvo |
| 51 | 58599 | *Herpetosiphon aurantiacus* | 6,79 | 23 | Hau |
| 52 | 58659 | *Salinispora arenicola* | 5,79 | 21 | Sar |
| 53 | 58565 | *Salinispora tropica* | 5,18 | 20 | Str |
| 54 | 58253 | *Bacteroides vulgatus* | 5,16 | 19 | Bvu |
| 55 | 57665 | *Deinococcus radiodurans R1* | 3,28 | 19 | DraR1 |
| 56 | 57879 | *Methanosarcina acetivorans C2A* | 5,75 | 18 | MacC2A |
| 57 | 58855 | *Sulfurihydrogenibium sp. YO3AOP1* | 1,84 | 17 | SYO3AOP1 |
| 58 | 57613 | *Bordetella bronchiseptica* | 5,34 | 15 | Bbr |
| 59 | 57885 | *Fusobacterium nucleatum* | 2,17 | 14 | Fnu |
| 60 | 57863 | *Ruegeria pomeroyi* | 4,59 | 13 | Rpo |
| 61 | 58095 | *Zymomonas mobilis* | 2,06 | 13 | Zmo |
| 62 | 57823 | *Burkholderia xenovorans LB400* | 9,74 | 9 | BxeLB400 |
| 63 | 58743 | *Shewanella baltica OS185* | 5,31 | 9 | SbaOS185 |
| 64 | 58775 | *Shewanella baltica OS223* | 5,36 | 6 | SbaOS223 |

Table S2. The list of 64 reference genomes for the SYNTH dataset ordered in the decreasing order of their coverage depths.

| No. | Species | NGA50 | | | | # intragenomic misassemblies | | | |
|---|---|---|---|---|---|---|---|---|---|
| | | metaSPAdes | MEGAHIT | IDBA-UD | Ray-Meta | metaSPAdes | MEGAHIT | IDBA-UD | Ray-Meta |
| 1 | Neq | 262512 | 474066 | 474066 | 474106 | 1 | 0 | 1 | 0 |
| 2 | Pho | 186786 | 114964 | 298215 | 296501 | 2 | 0 | 1 | 4 |
| 3 | Rba | 186846 | 113658 | 220154 | 159603 | 4 | 1 | 3 | 23 |
| 4 | ThRQ2 | 26177 | 3128 | 6960 | 13096 | 0 | 0 | 0 | 26 |
| 5 | Afu | 183804 | 82225 | 85088 | 163672 | 2 | 2 | 2 | 9 |
| 6 | Neu | 45789 | 45729 | 46450 | 46140 | 2 | 2 | 2 | 5 |
| 7 | ThDSM4359 | 52398 | 42518 | 54328 | 57472 | 0 | 0 | 0 | 4 |
| 8 | Sto | 76823 | 48033 | 58743 | 67066 | 0 | 0 | 0 | 6 |

| # | Name | | | | | | | | |
|---|---|---|---|---|---|---|---|---|---|
| 9 | HY04AAS1 | 399781 | 114387 | 129866 | 148210 | 0 | 0 | 0 | 4 |
| 10 | Gau | 1590303 | 345901 | 618807 | 230304 | 0 | 0 | 0 | 2 |
| 11 | PaeIM2 | 91533 | 57037 | 69473 | 106833 | 1 | 0 | 3 | 5 |
| 12 | Pfu | 59599 | 51223 | 54607 | 57958 | 0 | 0 | 0 | 9 |
| 13 | Cvi | 177554 | 134847 | 134659 | 224311 | 3 | 0 | 4 | 2 |
| 14 | Aca | 268175 | 142947 | 170627 | 131556 | 0 | 0 | 1 | 9 |
| 15 | Pca | 276263 | 126050 | 140648 | 194759 | 0 | 0 | 0 | 1 |
| 16 | Abo | 56874 | 86643 | 125033 | 45173 | 0 | 0 | 0 | 3 |
| 17 | GsuPCA | 188250 | 163216 | 187511 | 134801 | 4 | 4 | 4 | 11 |
| 18 | PmaEX-H1 | 1063166 | 1063325 | 549093 | 515028 | 0 | 0 | 0 | 0 |
| 19 | SEE-36 | 8209 | 1178 | 2338 | 2865 | 2 | 3 | 2 | 9 |
| 20 | SNAS-14.1 | 4005 | 1314 | 1993 | 2892 | 8 | 7 | 5 | 15 |
| 21 | Mja | 121749 | 57235 | 66977 | 101663 | 1 | 1 | 0 | 2 |
| 22 | Tde | 169913 | 73548 | 121352 | 120750 | 0 | 0 | 4 | 7 |
| 23 | Mka | 984861 | 223403 | 223403 | 562320 | 0 | 0 | 0 | 3 |
| 24 | Pas | 154757 | 127761 | 127087 | 132679 | 0 | 0 | 1 | 4 |
| 25 | SyeSS-5 | - | 1331 | - | 1185 | 4 | 3 | 16 | 3 |
| 26 | Cte | 149407 | 100902 | 128768 | 107579 | 0 | 0 | 2 | 1 |
| 27 | MmaC5 | 169017 | 22399 | 23198 | 48711 | 0 | 0 | 0 | 9 |
| 28 | Dtu | 938843 | 113442 | 178437 | 179329 | 0 | 0 | 0 | 0 |
| 29 | TpeRKU-1 | - | 3068 | 1990 | 3100 | 0 | 1 | 1 | 2 |
| 30 | TthHB8 | 60940 | 54274 | 58842 | 35334 | 1 | 0 | 0 | 3 |
| 31 | Cli | 104004 | 79065 | 101242 | 83504 | 1 | 1 | 4 | 4 |
| 32 | DesPig | 109658 | 89070 | 90236 | 38875 | 16 | 6 | 8 | 26 |
| 33 | Csa | 39297 | 25705 | 26050 | 35961 | 4 | 7 | 7 | 20 |
| 34 | Wsu | 156243 | 138697 | 138697 | 138917 | 0 | 0 | 0 | 0 |
| 35 | MmaS2 | 109893 | 22868 | 15651 | 85289 | 1 | 0 | 0 | 6 |
| 36 | Cph | 40887 | 38781 | 43588 | 39901 | 9 | 3 | 3 | 7 |
| 37 | Pph | 89288 | 76050 | 75302 | 56959 | 0 | 0 | 1 | 11 |
| 38 | CauJ-10-fl | 78373 | 46382 | 67634 | 30469 | 8 | 7 | 7 | 27 |
| 39 | Amu | 176763 | 107931 | 130111 | 90381 | 2 | 0 | 0 | 4 |
| 40 | Cth | 74152 | 53563 | 57019 | 54399 | 4 | 3 | 4 | 3 |
| 41 | Pgi | 30817 | 26754 | 29095 | 21559 | 3 | 2 | 5 | 6 |
| 42 | Efa | 50949 | 41132 | 41368 | 41681 | 50 | 47 | 49 | 50 |
| 43 | Cbe | 39468 | 26834 | 25903 | 38981 | 1 | 5 | 8 | 7 |
| 44 | Tps | 54527 | 48090 | 51478 | 32075 | 0 | 1 | 1 | 6 |
| 45 | Lch | 15446 | 15355 | 14870 | 3469 | 2 | 1 | 9 | 9 |
| 46 | NPCC7120 | 138267 | 79686 | 91348 | 27221 | 4 | 1 | 4 | 13 |
| 47 | DvuDP4 | 88453 | 80883 | 106219 | 15645 | 18 | 12 | 13 | 23 |
| 48 | Iho | 212224 | 78313 | 78313 | 23087 | 0 | 0 | 1 | 3 |
| 49 | Bth | 132888 | 108389 | 131935 | 26522 | 8 | 3 | 5 | 18 |
| 50 | Hvo | 22411 | 24160 | 22395 | 3467 | 0 | 0 | 0 | 2 |
| 51 | Hau | 112184 | 120299 | 139818 | 13979 | 4 | 4 | 2 | 15 |
| 52 | Sar | 10677 | 9693 | 8544 | 1994 | 3 | 5 | 5 | 2 |
| 53 | Str | 9852 | 8526 | 7698 | 1934 | 2 | 1 | 7 | 4 |
| 54 | Bvu | 88327 | 78679 | 78066 | 7488 | 1 | 3 | 8 | 9 |
| 55 | DraR1 | 15738 | 14961 | 15007 | 1649 | 0 | 0 | 1 | 2 |
| 56 | MacC2A | 26007 | 22323 | 24262 | 4846 | 9 | 8 | 9 | 12 |
| 57 | SYO3AOP1 | 6617 | 2016 | 6496 | 8143 | 2 | 9 | 1 | 38 |
| 58 | Bbr | 5634 | 5358 | 5074 | 1144 | 5 | 1 | 20 | 1 |
| 59 | Fnu | - | - | - | - | 0 | 0 | 0 | 1 |
| 60 | Rpo | 12719 | 12752 | 12979 | 1078 | 0 | 1 | 9 | 2 |
| 61 | Zmo | 33151 | 32449 | 42083 | 1359 | 1 | 1 | 1 | 1 |
| 62 | BxeLB400 | 4868 | 4425 | 4535 | - | 12 | 10 | 61 | 4 |
| 63 | SbaOS185 | 8588 | 2917 | 6341 | - | 6 | 7 | 6 | 1 |
| 64 | SbaOS223 | - | 1524 | - | - | 8 | 7 | 3 | 1 |

Table S3. NGA50 statistics and the number of intragenomic misassemblies for 64 reference genomes for the SYNTH dataset arranged in the decreasing order of their coverage depths. The colors of the cells reflect how much the results of various assemblers differ from the median value (blue/red cells indicate that the results improve/deteriorate as compared to the median value).

# Supplementary Text F: Analysis of the CAMI dataset

| No. | Taxonomic ID | Organism name | Genome size (Mbp) | Average coverage |
|---|---|---|---|---|
| 1 | 1247738.1 | *Campylobacter coli BIGS0015* | 1,3 | 257 |
| 2 | 1097667.1 | *Patulibacter medicamentivorans* | 4,77 | 200 |
| 3 | 1399144.1 | *Brevibacillus laterosporus PE36* | 5,11 | 199 |
| 4 | 494419.1 | *Arthrobacter sp. TB 23* | 3,47 | 166 |
| 5 | 314254.1 | *Oceanicaulis sp. HTCC2633* | 3,17 | 140 |
| 6 | 290399.1 | *Arthrobacter sp. FB24* | 5,07 | 137 |
| 7 | 883112.1 | *Facklamia ignava CCUG 37419* | 1,76 | 133 |
| 8 | 434085.1 | *gamma proteobacterium IMCC2047* | 0,46 | 133 |
| 9 | 1131272.1 | *Chloroflexi bacterium SCGC AB-629-P13* | 0,79 | 108 |
| 10 | 1224136.1 | *Enterobacteriaceae bacterium LSJC7* | 4,6 | 96 |
| 11 | 1123317.1 | *Streptococcus sobrinus DSM 20742 = ATCC 33478* | 1,74 | 89 |
| 12 | 457393.1 | *Bacteroides sp. 4_1_36* | 4,61 | 88 |
| 13 | 1353530.1 | *Bacteriovorax sp. DB6_IX* | 2,51 | 82 |
| 14 | 1159204.1 | *Mycoplasma gallisepticum NC08_2008.031-4-3P* | 0,93 | 79 |
| 15 | 1209372.1 | *Bacillus sp. WBUNB009* | 5,58 | 77 |
| 16 | 1263006.1 | *Firmicutes bacterium CAG:170* | 2,27 | 77 |
| 17 | 1386080.1 | *Bacillus sp. EGD-AK10* | 4,33 | 76 |
| 18 | 1386078.1 | *Pseudomonas sp. EGD-AK9* | 3,88 | 70 |
| 19 | 322710.1 | *Azotobacter vinelandii DJ* | 5,37 | 68 |
| 20 | 766138.1 | *Shigella boydii 965-58* | 5,15 | 59 |

Table S4. The list of 20 most abundant reference genomes in the CAMI dataset arranged in the in decreasing order of their coverage depths.

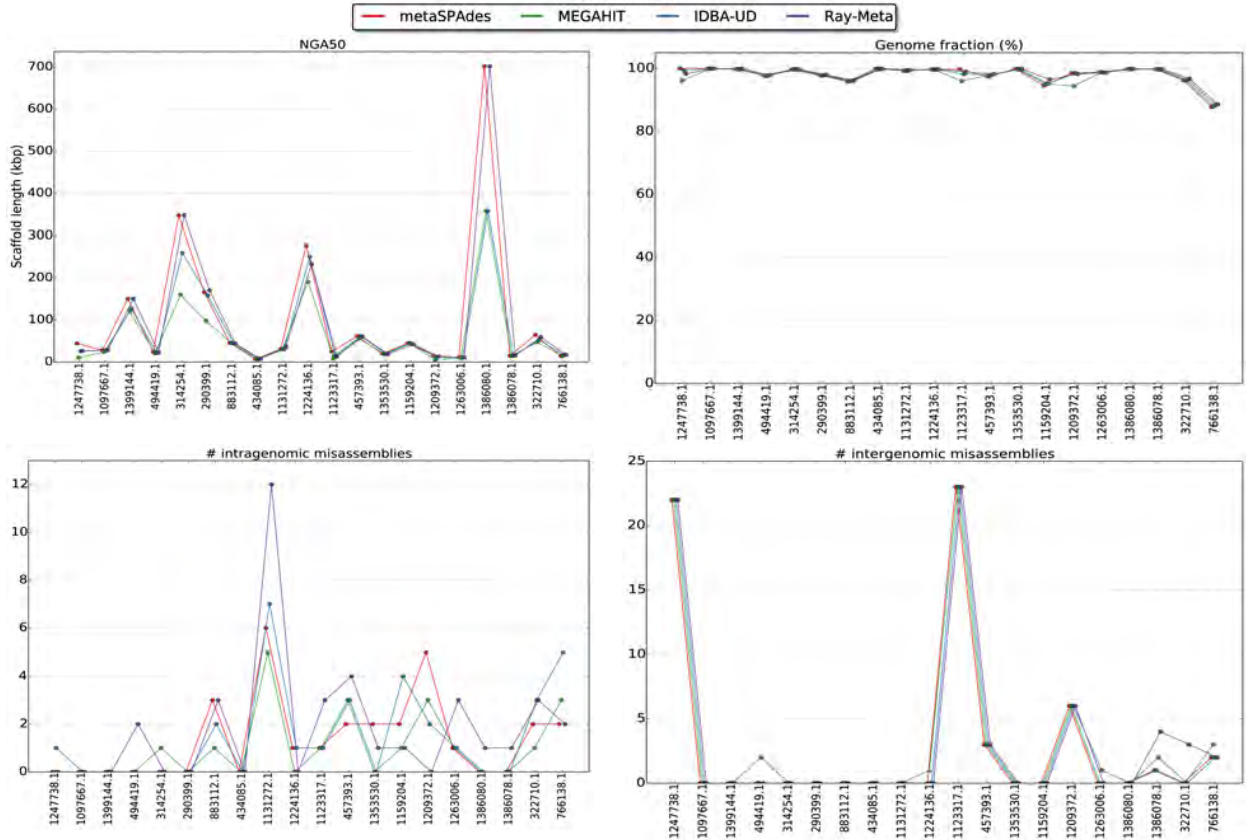

Figure S2. The NGA50 statistics (top left), the fraction of the reconstructed genome (top right) the number of intragenomic misassemblies (bottom left) and the number of intergenomic misassemblies (bottom right) for 20 most abundant species from the CAMI dataset. The genomes are arranged in the decreasing order of their coverage depths.

In addition to the CAMI dataset described in the main text, we also analyzed a lower complexity dataset generated by the CAMI consortium (referred to as $CAMI_{low}$), which was simulated from 30 reference genomes (Table S5). The results of benchmarking on the $CAMI_{low}$ dataset are summarized in Table S6, Figure S3 and Figure S4.

| No. | Taxonomic ID | Organism name | Genome size (Mbp) | Average coverage |
|---|---|---|---|---|
| 1 | 434085.1 | *Gamma proteobacterium IMCC2047* | 2,23 | 873 |
| 2 | 247639.1 | *Marine gamma proteobacterium HTCC2080* | 3,58 | 53 |
| 3 | 1050222.1 | *Paenibacillus sp. Aloe-11* | 5,81 | 22 |
| 4 | 667138.1 | *Thermoplasmatales archaeon I-plasma* | 1,69 | 21 |
| 5 | 552396.1 | *Erysipelotrichaceae bacterium 5_2_54FAA* | 6,26 | 16 |
| 6 | 1007115.1 | *Gamma proteobacterium SCGC AAA076-D13* | 1,66 | 14 |
| 7 | 1122939.1 | *Patulibacter americanus DSM 16676* | 4,47 | 9 |
| 8 | 1111069.1 | *Thermus sp. CCB_US3_UF1* | 2,26 | 8 |
| 9 | 1131272.1 | *Chloroflexi bacterium SCGC AB-629-P13* | 0,84 | 8 |
| 10 | 1131273.1 | *Marinimicrobia bacterium SCGC AB-629-J13* | 1,93 | 8 |
| 11 | 1097667.1 | *Patulibacter medicamentivorans* | 5,09 | 7 |
| 12 | 1263001.1 | *Firmicutes bacterium CAG:114* | 2,34 | 4 |
| 13 | 1137281.1 | *Formosa sp. AK20* | 3,06 | 3 |
| 14 | 1345697.1 | *Geobacillus sp. JF8* | 3,49 | 2 |
| 15 | 1412874.1 | *Uncultured archaeon A07HR60* | 2,88 | 1,9 |
| 16 | 1224136.1 | *Enterobacteriaceae bacterium LSJC7* | 4,61 | 1,8 |
| 17 | 1229484.1 | *Alpha proteobacterium LLX12A* | 5,96 | 1,4 |
| 18 | 1229781.1 | *Brevibacterium casei S18* | 3,66 | 1,2 |
| 19 | 1235799.1 | *Lachnospiraceae bacterium 3-2* | 4,46 | 1,0 |
| 20 | 370895.1 | *Burkholderia mallei 2002721280* | 5,68 | 0,9 |
| 21 | 742723.1 | *Lachnospiraceae bacterium 2_1_46FAA* | 4,43 | 0,9 |
| 22 | 1045854.1 | *Weissella koreensis KACC 15510* | 1,44 | 0,7 |
| 23 | 1009708.1 | *Alpha proteobacterium SCGC AAA536-G10* | 2,16 | 0,6 |
| 24 | 1174684.1 | *Sphingopyxis sp. MC1* | 3,65 | 0,4 |
| 25 | 349101.1 | *Rhodobacter sphaeroides ATCC 17029* | 4,49 | 0,4 |
| 26 | 1230476.1 | *Bradyrhizobium sp. DFCI-1* | 7,65 | 0,3 |
| 27 | 245012.1 | *Butyrate-producing bacterium SM4/1* | 3,11 | 0,3 |
| 28 | 939301.1 | *Alpha proteobacterium SCGC AAA015-O19* | 1,74 | 0,2 |
| 29 | 1263006.1 | *Firmicutes bacterium CAG:170* | 2,45 | 0,2 |
| 30 | 1394711.1 | *Candidatus Saccharibacteria bacterium RAAC3_TM7_1* | 0,85 | 0,1 |

Table S5. The list of 30 reference genomes comprising the CAMI$_{low}$ dataset arranged in the decreasing order of their coverage depths.

| dataset/ assembler | metaSPAdes | | | MEGAHIT | | | IDBA-UD | | | Ray-Meta | | |
|---|---|---|---|---|---|---|---|---|---|---|---|---|
| | 10 | 1000 | ALL | 10 | 1000 | ALL | 10 | 1000 | ALL | 10 | 1000 | ALL |
| CAMI$_{low}$ | **5.5** | 40.3 | 65.2 | 4.8 | 40.4 | 65.4 | 4.7 | **41.9** | **66.3** | 4.5 | 33.6 | 42.0 |

Table S6. The total length of scaffolds generated by metaSPAdes, MEGAHIT, IDBA-UD, and Ray-Meta (in megabases) for CAMI$_{low}$ dataset. Statistics are shown for 10 longest, 1000 longest and all scaffolds longer than 1 kb. The top results among all assemblers are highlighted in bold.

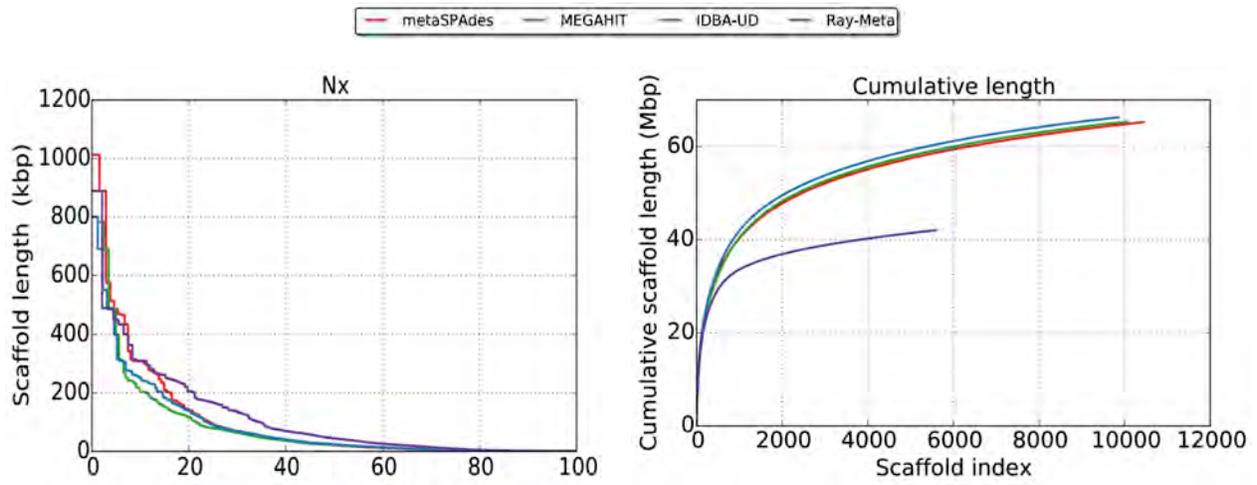

Figure S3. Nx plot (left) and the cumulative scaffold length plot (right) for CAMI$_{low}$ dataset.

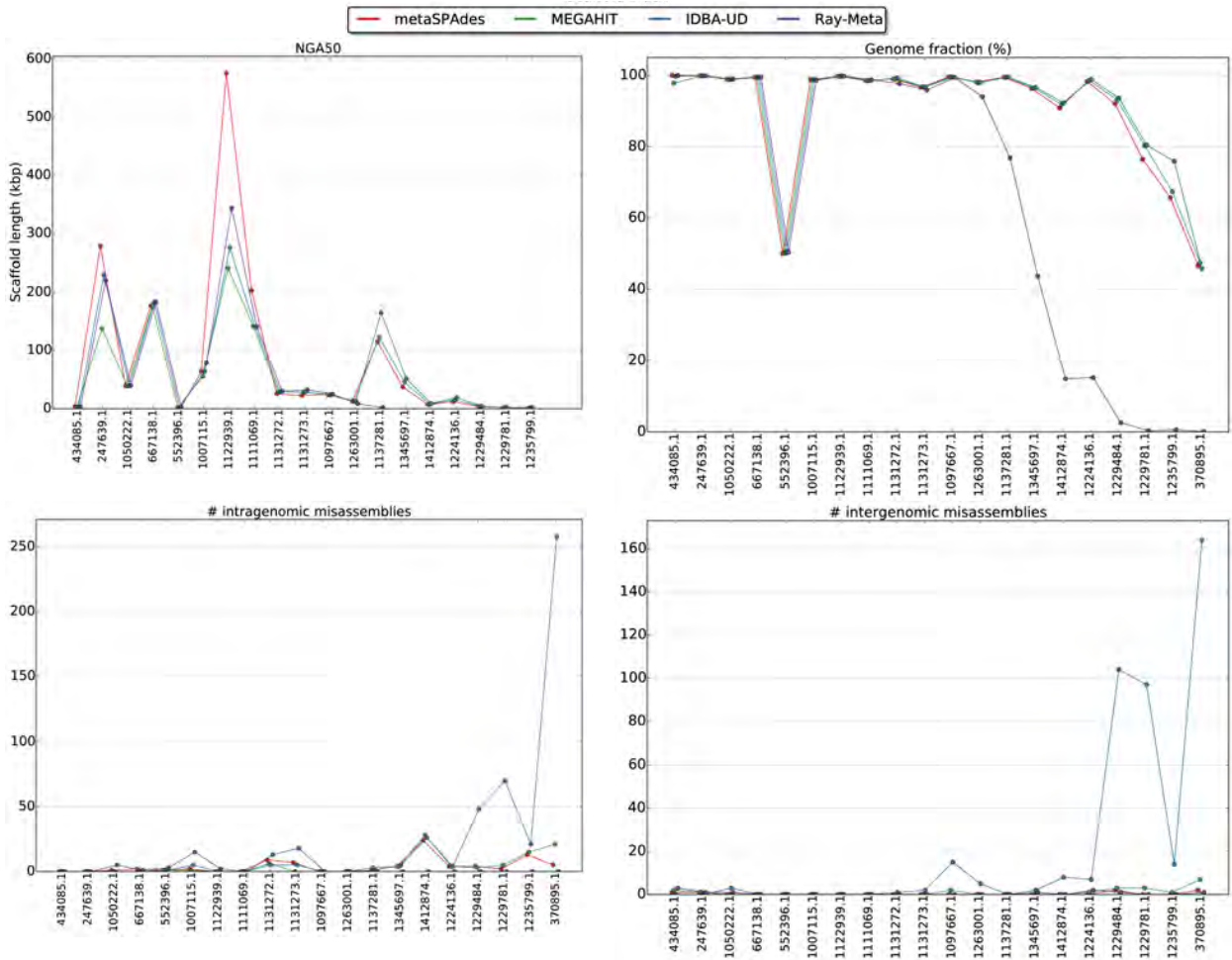

Figure S4. The NGA50 statistics (top left), the fraction of the reconstructed genome (top right), the number of intragenomic misassemblies (bottom left) and the number of intergenomic misassemblies (bottom right) for 20 most abundant species comprising CAMI$_{low}$ dataset. References are specified by their Taxonomic IDs (see Table S5) and arranged in the decreasing order of their coverage depths.

**Supplementary text G: Analysis of the HMP dataset**

As described in the main text, we identified only three references that were at least 70% covered by contigs generated by at least one of four assemblers analyzed in this study (*Streptococcus salivarius SK126, Neisseria subflava NJ9703*, and *Prevotella melaninogenica ATCC 25845* abbreviated as *Ssa, Nsu, and Pme,* respectively). Figure S5 presents benchmarking results for these three genomes.

Note that the number of reported errors in the HMP assembly (Figure S5, bottom) significantly exceeds the number of errors for SYNTH and CAMI datasets or the number of errors in typical assemblies of isolates. We believe that most of these errors represent metaQUAST artifacts (rather than true assembly errors) caused by the significant differences between the three recruited references and the related genomes in the sample. Poor coverage of the two out of three references genomes by the assembly contigs also suggests that these differences contributed to the metaQUAST artifacts.

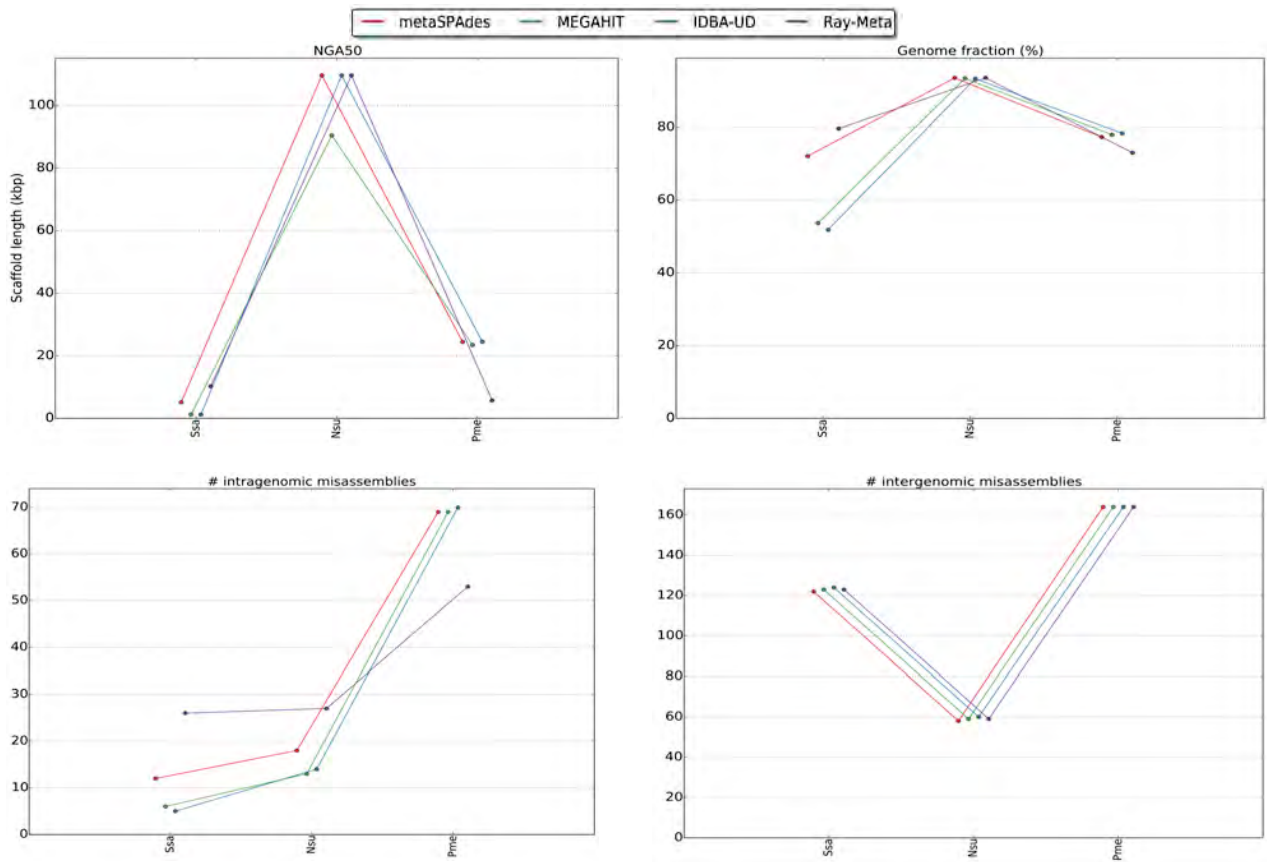

Figure S5. The NGA50 statistics (top left), the fraction of the reconstructed genome (top right), the number of intragenomic misassemblies (bottom left) and the number of intergenomic misassemblies (bottom right) for three reference genomes identified for the HMP dataset. References are placed in the decreasing order of their average coverage-depths (183X, 118X, and 15X for *Ssa, Nsu*, and *Pme,* respectively).